# Low Polarization Sensitive O-band SOA on InP Membrane for Advanced Photonic Integration

Desalegn Wolde Feyisa, *Student Member, IEEE*, Salim Abdi, *Student Member, IEEE*, Rene van Veldhoven, Nicola Calabretta, *Member, IEEE,* Yuqing Jiao, *Senior Member, IEEE*, and Ripalta Stabile, *Senior Member, IEEE*

*Abstract*—Managing insertion losses, polarizations and device footprint is crucial in developing large-scale photonic integrated circuits (PICs). This paper presents a solution to these critical challenges by designing a semiconductor optical amplifier (SOA) in the O-band with reduced polarization sensitivity, leveraging the ultra-compact InP Membrane on Silicon (IMOS) platform. The platform is compatible with close integration atop electronics, via densely populated vertical interconnects. The SOA incorporates a thin tensile-strained bulk active layer to mitigate polarization sensitivity. The developed 500 µm long SOA has a peak gain of 11.5 dB at 1350 nm and an optimal polarization dependency of less than 1 dB across a 25 nm bandwidth, ranging from 1312 nm to 1337 nm. The device is practical for integrated circuits where multiple amplifiers work in cascades with a minimal 6.5 dB noise figure (NF) measured at the gain peak. The designed vertical active-passive transition, achieved through inverse tapering, allows for effective field coupling in the vertical direction resulting in a transmission efficiency of over 95% at the transition and minimal polarization sensitivity of less than 3%. The device yields significant gain at a small current density of less than 3 kA/cm² as the result of minimalist gain medium structure, reducing joule heating and improving energy efficiency. This is especially relevant in applications such as optical switching, where multiple SOAs populate the PIC within a small area. Consequently, the simulated and fabricated low polarization sensitive O-band SOA is a suitable candidate for integration into large-scale, ultra-compact photonic integrated circuits.

*Index Terms*—Active passive transition, InP membrane on Silicon, Low noise figure, Photonic integrated circuits, polarization sensitivity, semiconductor optical amplifiers.

## I. Introduction

PHOTONIC Integrated Circuits (PICs) are vital in advancing optical solutions for next-generation communication and computing technologies. PICs have already demonstrated their potential impact in practical use cases, including optical switch matrices, neuromorphic computing, and network on chip [1]–[4]. However, key challenges persist in this domain, including polarization handling, insertion losses arising from components and facet coupling, the substantial size of the components, and the complexities associated with achieving large-scale integration with seamless active-passive transitions.

The inclusion of gain devices like Erbium doped waveguide amplifiers (EDWA) and SOAs can effectively compensate for signal loss. SOAs, however, offer additional advantages such as compactness, cost-efficiency, broad gain bandwidth, and adaptability for large-scale monolithic photonic integrated circuits, making them well-suited for advanced optical signal processing [5] . Furthermore, having compact and robust passive devices co-integrated with SOA is imperative for fabricating PICs self-sustained in power budget while having a compact footprint, a feat that is realizable within the InP membrane on silicon (IMOS) platform. The IMOS platform stands out primarily for its compact passive components due to its intrinsic high refractive index and the availability of native active devices [6], [7]. While other high-index contrast platforms, like Silicon on Insulator (SOI), similarly offer compact passive components, the co-integration of the amplifier predominantly necessitates a more challenging process of heterogeneous integration, and polarization insensitive (PI) SOA has not been demonstrated [8]. Additionally, the ability of IMOS to be integrated vertically with electronics enables highly dense interconnects, offering significant advantages for developing large-scale, high-speed, and complex optoelectronic circuitries [9]. However, polarization dependent amplification notably arises from the asymmetry of the gain medium waveguide and the transition sections between active and passive components, which makes polarization independent (PI) PICs still a subject of ongoing research.

Prior research endeavors have explored the reduction of polarization sensitivity by incorporating supplementary on-chip polarization management components or employing techniques that change the structural configuration of the SOA. The former approach necessitates the inclusion of an auxiliary passive polarization rotator along with two congruent SOA structures [10], [11], which increases the complexity of the device. The latter method is favored because it only requires modifications of the SOA structure, resulting in a smaller number of

This work was supported in part by H2020 ICT TWILIGHT project (Contract No. 781471) under the Photonics Public Private Partnership. *Desalegn Wolde Feyisa and Salim Abdi contributed equally*, Corresponding author: Desalegn Wolde Feyisa.)

Desalegn Wolde Feyisa, Salim Abdi, Nicola Calabretta, Yuqing Jiao and Ripalta Stabile are with Eindhoven Hendrik Casimir Institute (EHCI), Eindhoven University of Technology, Eindhoven 5600MB, the Netherlands (e-mail: d.w.feyisa@tue.nl; s.a.abdi@tue.nl; n.calabretta@tue.nl; y.jiyao@tue.nl; r.stabile@tue.nl )

René van Veldhoven, is with Nanolab@TU/e, Eindhoven University of Technology, Eindhoven 5600MB, the Netherland (e-mail: p.j.veldhoven@tue.nl ).

components and component interfaces. Consequently, this approach helps minimize losses related to components and interface reflections.

Earlier studies have demonstrated the realization of PI SOA on alternative platforms utilizing strained Multiple Quantum Wells (MQW) [12], square-shaped active layer by employing thick active layer bulk [13], [14], and tensile strained active bulk layer [15]. All these approaches present challenges. Within quantum wells, uniformly minimizing polarization sensitivity across all wavelengths and operating currents presents substantial challenges, due to the pronounced dependency of material gain on both wavelength and current. Employing a thick bulk-active layer strategy enables comparable confinement factors for both polarizations. Nonetheless, the increased thickness of the active layer leads to an increase in transparency current and less effective heat management, becoming the limiting factor in dense integration strategies and membrane photonics. The most feasible way is to use thin active layers and introduce tensile strain to increase TM material gain to compensate for the low confinement factor stemming from the structural asymmetry. Such thin active layers need reduced driving currents compared to their thicker counterparts. Furthermore, advancements in semiconductor processing technologies have enabled more accurate fabrication of epitaxial layer stacks in devices, allowing for the attainment of desired strain and layer thickness with considerable precision.

This paper presents the design, fabrication, and evaluation of an O-band (to leverage legacy O-band networks in data center) PI SOA within the IMOS platform, using tensile-strained quaternary InGaAsP bulk material as the active layer and improving thermal dissipation. Section II shows the design aspects of the layer stack and the integration of active-passive components for high gain and efficient transmission and introduces a novel heat sink design for heat management. Section III details the layer growth, the characterization of its quality and the device fabrication, outlining the realization approach on the membrane platform. Section IV shows experimental results of transparency current, SOA gain, polarization-dependent gain (PDG), noise figure (NF), and optical signal to noise ratio (OSNR). Finally, Section V summarizes the findings, discusses potential improvements, and suggests implications of the exploitation of the developed SOAs in emerging optical networks.

## II. DESIGN OF LAYER STACK AND ACTIVE-PASSIVE TRANSITION

This section details three pivotal components in the design phase. First, it presents the conceptualization of the layer stack, aiming to achieve polarization insensitivity (sub-section II-A). Following this, it introduces the design of the heat sink, illustrating how it directs heat from the membrane diode structure to the silicon carrier (sub-section II-B). Finally, it details the active-passive transition, where light couples from the 2 μm wide SOA structure to the 330 nm wide InP waveguide, with minimal compromise of the polarization insensitivity and coupling efficiency (sub-section II-C).

### A. Layer Stack Design for the O-band PI SOA

In this study, we have designed and simulated an InGaAsP/InP-based SOA using the HAROLD commercial active device layer stack simulator from Photon Design, which can solve the Poisson, the current continuity, the capture/escape balance, and the photon rate equations. The designed target for photoluminescence (PL) peak wavelength is at 1350 nm since the band-filling effect is expected to cause a peak blue shift to 1310 nm when the SOAs operate at 4 kA/cm$^2$, assuming we can neglect heating effects in the case of an effective heatsink. Fig. 1 presents the complete epitaxial layer stack, as defined in Harold and Fig. 2a shows the final SOA structure as it appears after fabrication. Here, the active layer, Q* (layer 5), composed of a tensile strained (0.18%) bulk quaternary material with a bandgap wavelength of 1.35 μm, is positioned between two separate confinement heterostructure (SCH) layers with a bandgap wavelength of 1.05 μm (layers 4 and 6). The selected bandgap of the SCH ensures that the band offset with the active core guarantees adequate electrical confinement. The cumulative thickness of the active region is 300 nm, which includes the 50 nm active layer and 125 nm SCH on both the top and the bottom, respectively. This active waveguide thickness successfully guarantees effective single mode guidance. The waveguiding core is buried between n- and p-type layers. The 800 nm InP p-cladding layer separates the highly p-doped InGaAs layer from the mode propagation region. On the lower end, the 300 nm InP layer is used for passive functionalities, while the 300 nm InGaAs layer functions as an etch stop during the wet etching process of the InP wafer after bonding. The thickness of the active layer is kept well above the threshold of quantum confinement to ensure bulk behavior and provide sufficient optical confinement factor. Finite difference element mode simulations are employed on the ultimate SOA structure depicted in Fig. 2a to guide the fundamental modes in the active region, as illustrated in Fig. 2b. These simulations reveal an optical confinement of approximately 11% for the fundamental Transverse Electric mode (TE) and 8.5% for the fundamental Transverse Magnetic mode (TM) within the active region for a 2 μm wide s-shaped

| Layer # | Layer | Material | Thickness(nm) | Doping(cm$^{-3}$) |
|---|---|---|---|---|
| 1 | cap layer | InP | 50 | nid |
| 2 | contact | InGaAs | 50 | p=8×10$^{18}$ |
| 3 | contact | InP | 800 | p=1×10$^{18}$ |
| 4 | SCH | Q1.05 | 125 | nid |
| 5 | Bulk active | Q* | 50 | nid |
| 6 | SCH | Q1.05 | 125 | nid |
| 7 | contact | InP | 80 | n=4×10$^{18}$ |
| 8 | Waveguide | InP | 300 | nid |
| 9 | etch stop | InGaAs | 300 | nid |
| 10 | substrate | InP | | n=1×10$^{18}$ |

Figure 1. Epitaxial layer stack defined with Harold. Q* represents Quaternary InGaAsP material with 0.18% tensile strain.

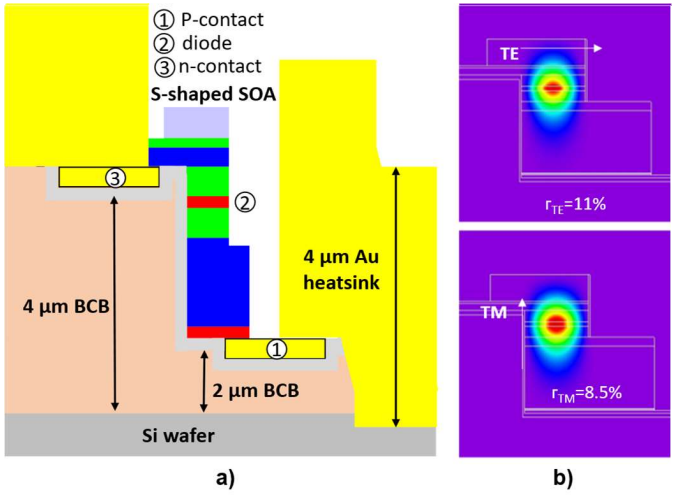

Figure. 2. a) schematic illustration of final cross-section of the SOA with heat sink b) TE and TM mode as simulated from final device schematic in fig a structure. This confinement is comparable to the confinement factor of 8 quantum-well, as previously demonstrated for this same platform [16]. The internal SOA gain in relation to the confinement factor $\Gamma$ and material gain $g_m$ is given as:

$$G = 10 \log_{10}(e^{(\Gamma g_m - \alpha)l}), \quad 1$$

where G is the internal SOA gain and α is the material loss.

The design target is to minimize the PDG, calculated as the difference between the TE ~~modal gain~~ internal SOA gain ($G_{TE}$) and the TM ~~modal~~ internal SOA gain ($G_{TM}$), which in turn are related to the confinement factor and the material gain, as shown in the following equations, while neglecting polarization dependent material losses:

$$PDG = |G_{TE} - G_{TM}|$$
$$= 10 \log_{10}(e^{(\Gamma_{TE}g_{TE} - \Gamma_{TM}g_{TM})l}),$$
$$= 10 \log_{10} e^{(1-\Gamma_{ratio}g_{ratio})g_{TE}}, \quad 2$$

where $g_{ratio}$ is the ratio of material gains ($g_{TM}/g_{TE}$), $g_{TM}$ is TM material gain, $g_{TE}$ is TE material gain, $\Gamma_{ratio}$ is the confinement factor ratio ($\Gamma_{TM}/\Gamma_{TE}$). The PDG depends on the product of the confinement factor ratio by the material gain ratio (neglecting polarization dependent material carrier absorption losses). Achieving ideal polarization insensitivity is possible when this product equals unity. However, the asymmetric waveguide structure and material gain difference usually make this product smaller than 1, resulting in significant PDG. That is the case of MQW structures, which have anisotropic material gain (usually $g_{TM} < g_{TE}$) and anisotropic confinement factor (usually $\Gamma_{TM} < \Gamma_{TE}$) [17]. On the contrary, a bulk gain material has isotropic material gain ($g_{TM} = g_{TE}$) and the modal gain ($\Gamma \times g$) relates to the confinement factor and the material gain. Moreover, the epitaxial layer strain can be used to compensate for the confinement factor discrepancy to achieve polarization insensitivity. In fact, the principal contribution to the TE material gain originates from the stimulated transition occurring between the conduction band and the heavy hole (HH), while the TM material gain derives from the stimulated transition between the conduction band and the light hole (LH): Introducing tensile strain in the active region allows for the upward shifting of the LH band, subsequently enhancing TM material gain [18]. To realize PI gain conditions, we introduced a tensile strain of approximately 0.18% to the active region of the designed PI SOA device.

Fig. 3a and 3b show material gains for the TE and TM modes, respectively, as calculated from the designed strained layer stack. The material gain is shown for current values varying from 5 mA to 40 mA with step of 5 mA in the direction indicated by the blue arrow. The gain peak is at 1350 nm for a low current operation, while, at an operating current of 4 kA/cm$^2$, the peak gain shifts to about 1310 nm, showing around 40 nm blue shift due to the band filling effect. We can also observe that the TM material gain is higher than the TE material gains due to the tensile strain added in the active region, effectively compensating for the lower confinement factor of TM light within the SOA.

One final parameter that needs to be defined is the SOA width. The SOA width is a critical design parameter as the output saturation power depends on the product of the effective modal cross-section area, A=$\left(\frac{dw}{\Gamma}\right)$ and on the saturation intensity, I=$\left(\frac{hv}{\delta\tau}\right)$ [19], [20], as:

$$P_s = C * \left(\frac{dw}{\Gamma}\right) * \left(\frac{hv}{\delta\tau}\right), \quad 3$$

where C is the fiber to chip coupling efficiency, d is the active layer thickness, w is the device width, h is the Plank's constant, v is the frequency, δ is differential gain, and τ is the carrier lifetime. After fixing the thickness of the active layer, the SOA ~~width~~width, and its confinement factor, Γ, primarily influence the effective cross-sectional area. Fig. 3c shows 2 µm wide and 500 µm long SOA gain at a current density of 4 kA/cm$^2$ as a function of wavelength, simulated from the structure in Fig. 2a using PICWAVE, a commercial tool (from Photon Design) used for calculating the field evolution along propagation axis via ~~slow-varying~~slow varying envelop approximation. We can see that the modal gains for TE and TM modes can achieve polarization insensitivity as the tensile strain overcomes the disparity in confinement factor and for a large bandwidth. Fig. 3d shows gain versus output power for TE and TM modes at the same current density of 4 kA/cm$^2$ for SOA widths of 1.5 µm and 2.5 µm. The output saturation power is 11 dBm for 2.5 µm wide SOA, 2 dB more than the 1.5 µm wide SOA. Even though we can choose SOA width as low as 1 µm for effective confinement of both TE and TM modes, for the sake of achieving good output saturation power, we have chosen to operate with 2 µm wide SOAs in this work.

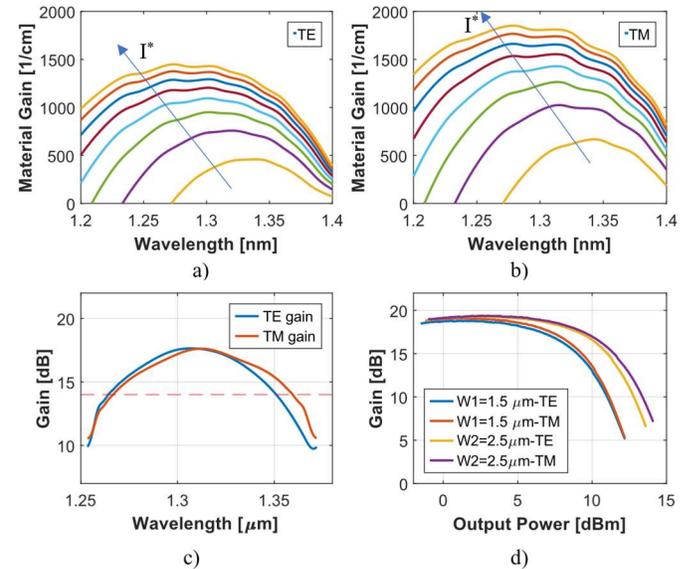

Figure. 3. a) Material gain for TE b) Material gain for TM c) ~~Modal G~~gain versus wavelength at current density 4kA/cm$^2$ d) ~~Modal gain~~Net gain versus output power for 500 μm long SOA at 4kA/cm$^2$ simulated for SOA width of 1.5 μm and 2.5 μm. I* represents current values from 5 mA to 40 mA with step of 5 mA per line in the direction indicated by the blue arrow.

*B. Heat Sink Design*

Bulk gain medium requires efficient heat sinking to prevent early gain saturation due to thermal effects. The schematic cross-sectional illustration of the s-shaped SOA with an integrated heat sink is shown in Fig. 2a. An Electroplated gold layer thicker than 4 µm is achieved to dissipate the heat towards the Silicon substrate. Here, a sufficient gap separates the heat sink from the diode structure to avoid optical signal absorption by the lossy gold metals. Simulations on the effect of <1 µm-thick gold on heat dissipation were reported in [21]. ~~Reduction of at least 23 °C is achieved with 200 nm-thick gold.~~ Furthermore, a reduction of at least 23 °C is achieved with 200 nm-thick gold. As COMSOL simulation suggests, by i~~I~~mplementing thick-plated gold to directly connect the diode p-contact to the silicon carrier further reduces the SOA ~~structure~~ temperature by 7-10°C for~~, i.e.,~~ a total reduction of 30-33°C.

*C. Polarization Insensitive Active Passive Transition*

Another crucial design aspect requires integrating active devices, like amplifiers, with passive components like waveguides, ensuring PI and low-reflection active-passive transition. One of the main challenges of monolithic integrations is to engineer the active and the passive components to ensure the best coupling and the lowest amount of interface reflections while ensuring low polarization sensitivity in the transition. The reflections can lead to ripple factors in the transmission spectrum, degrading SOA performance. To this end, a square-shaped O-band passive waveguide is designed with an optimal thickness of 300 nm and a width of 330 nm. These dimensions guarantee that the waveguide optimally supports and transmits fundamental TE and TM modes. Then, the waveguide is co-integrated with an active device utilizing the twin-guide approach, effectively reducing reflection, and achieving good coupling [22]. The SOA structure is laterally tapered as shown in Fig. 4 ~~(top inset)~~b. This decreases the effective refractive index in the SOA structure and, consequently, pushes the light down to the waveguide, as the optical light is naturally confined in the high effective refractive index part of the structure. Optimizing the taper length for each section in tandem with the taper tip allows for the attainment of the optimized coupling efficiency and reduced polarization sensitivity. Furthermore, the choice of the taper width is crucial for the PI transition. Fig. 4a illustrates the coupling efficiency from SOA structure to waveguide, simulated at taper section lengths of L1=20 µm, L2=20 µm, and L3=10 µm, as a function of taper width. Achieving a coupling efficiency of >95% for TM and 98% for TE is possible with a tapering width of <100 nm, satisfying our requirements. A taper width of 100 nm is selected as it offers adequate efficiency and low polarization dependence coupling without jeopardizing manufacturability. Although reducing taper width could elevate efficiency and mitigate polarization sensitivity, it lacks structural stability during fabrication. The light transition from top to bottom is observable in Fig. 4b (bottom ~~insets~~pictures). The TE field transitions from the active region to the waveguide before the TM field does. This occurs because the electric field for the TM field has a vertical orientation, meaning the lateral tapering impacts the TE~~M~~ field ~~less~~ more than the TM~~E~~ field, demonstrating that TE demands less stringent taper requirements than TM.

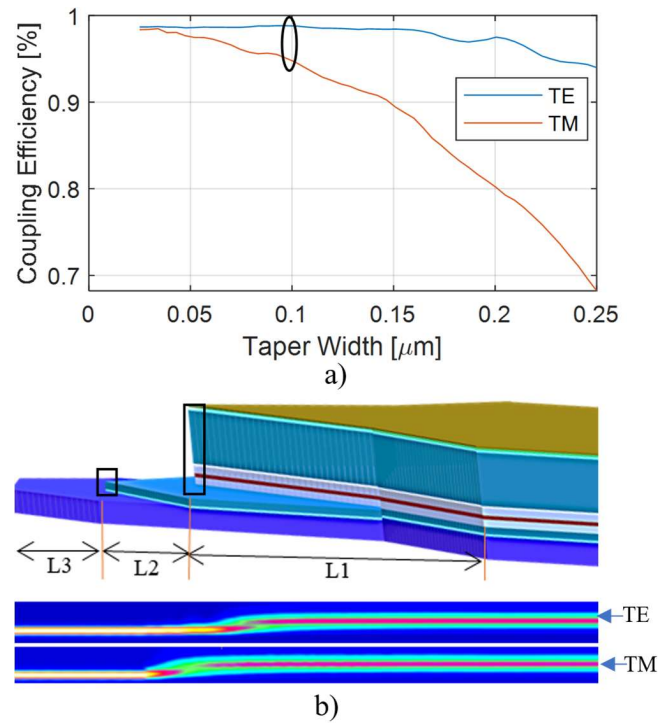

Figure. 4. Twin-guide active passive transition taper tip width effect on coupling efficiency and polarization sensitivity. b) Simplified 3D view of the active-passive transition together with TE and TM transitions. The device shown in this diagram is placed upside down in the final bonded membrane device.

III. SOA LAYER STACK GROWTH AND DEVICE FABRICATION

Before initiating the fabrication process, two tests were conducted on the developed layer stack to ascertain its quality. The initial step involved validating the accuracy of the emission wavelength peak through a photoluminescence (PL) test. Subsequently, a Transmission Electron Microscopy (TEM) analysis was employed to detect any defects in the active tensile-strained layer that might adversely affect mode propagation and amplification. Upon verification of the layer quality, the fabrication process commenced. The following subsections provide a detailed account of the testing procedures, steps involved in fabrication, and the observations recorded therein.

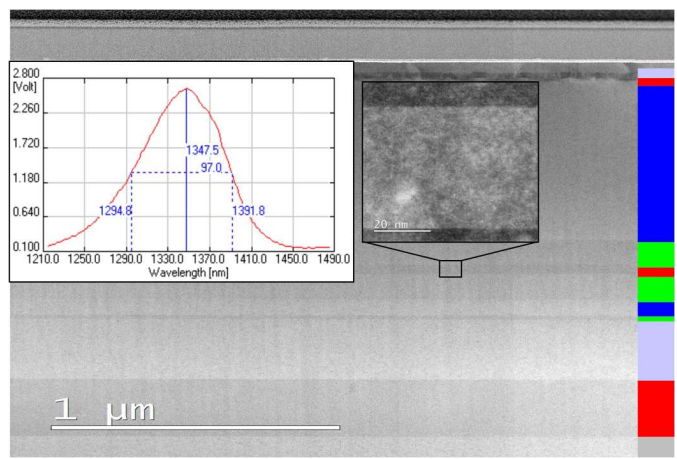

Figure. 5 TEM image of the stack, left right inset: atomic-resolution image of the active Quaternary (Q), right left inset: PL emission spectra of the active QQuaternary material.

*A. SOA layer stack growth and quality characterization*

The epitaxial growth is realized using low-pressure metalorganic vapor phase epitaxy (LP-MOVPE) on 3-inch n-type (100) InP wafers. The epitaxial stack is shown in Fig. 1 in section II. The crystallographic properties, along with the optical properties of the grown stack, are comprehensively characterized by high-resolution scanning transmission electron microscopy (HR-STEM), room-temperature photoluminescence (RTPL), and X-ray diffraction (XRD) measurements. Fig. 5 captures HR-STEM images obtained along the <011> zone axis, portraying the overall thickness and the active layer (right inset). The left inset corresponds to the active layer PL spectrum. It is worth noting that the 0.18 % strain accounted for in this study does not factor in the elastic strain reduction after the definition of mesa sidewalls [23]. Furthermore, the bonding process introduces a minimal additional tensile strain, approximately 0.03%, stemming from the coefficient of thermal expansion mismatch between InP and Si during bonding [24]. However, this, too, is not considered in the designed epitaxial strain values of 0.18%. More importantly, we note that the strained active bulk thickness of 50 nm with strain-thickness product of 9 %.nm is well below the critical thickness for relaxation via non-radiative defect formation [24]. The composition of the Quaternary (Q) active layer is calibrated within 1% tolerance in atomic composition to the design values to achieve 0.18% tensile strain and emission at 1350 nm, where 1% variation of Ga composition corresponds to 20 nm in wavelength shift, for instance. The HR-TEM image of the active core (right inset of Fig. 5) shows no defects and a lattice-matched interface between the active Q and SCH layer. Energy Dispersive X-ray Spectroscopy mapping of the active regions confirms the compositional uniformity of the layer with an average composition of 37.3%, 20.3%, 27.6 %, and 14.8% for In, Ga, As, and P elements, respectively. The full width at half maximum (FWHM) of the PL peak and its position (left inset of Fig. 5) also match the design values discussed in Section II.

*B. Device Fabrication process*

The fabrication process is shown in Fig. 6 and aAn image of the fabricated 3-inch InP membrane wafer is shown in Fig. 7 and the fabrication process is shown in Fig. 6. The twin-guide fabrication approach is realized on both sides of the InP wafer, where the bottom side is accessed after wafer-scale bonding onto Si. Unless specified otherwise, all lithography steps are done with e-beam lithography (EBL). The pattern is transferred to a plasma-enhanced chemical vapor deposition (PECVD) SiNx hard mask for all semiconductor etching steps. III-V semiconductor dry etching is done with CH4/H2 plasma, while wet etching uses HCl/H3PO4 and H2SO4/H2O2/H2O for etching InP and InGaAs/Q layers, respectively.

The process flow with active-passive integration is shown in Fig. 6 in 11 steps including:

1. Dry etching of the twin-guide first active-passive taper along with the n-mesa sidewall
2. Wet etching of the second taper into the n-cladding. The n-layers are also etched in the same step.
3. A 150 nm PECVD $SiO_2$ layer is deposited for surface passivation and opened where contacts are defined using a dry pure $CHF_3$ plasma process.
4. Then, n- and p-contacts are defined by e-beam evaporation and lift-off of Ni/Ge/Au and Ti/Pt/Au, respectively. The structures are then rapidly thermally annealed at 400°C for 30 seconds to achieve a low specific contact resistance. (Fig. 6a)

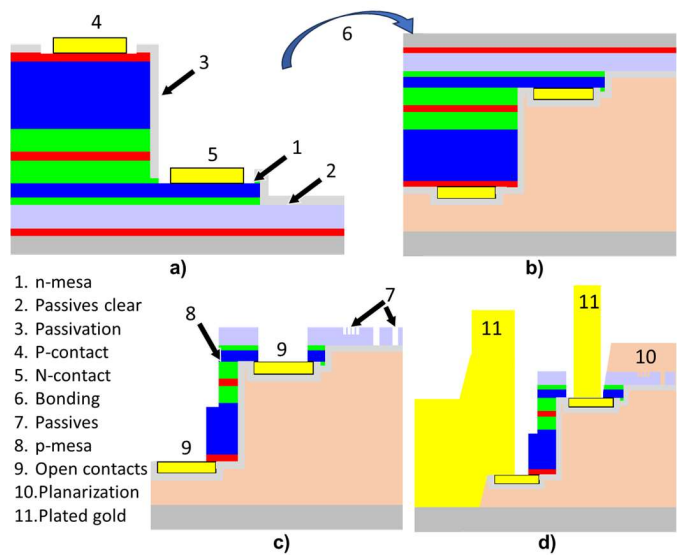

1. n-mesa
2. Passives clear
3. Passivation
4. P-contact
5. N-contact
6. Bonding
7. Passives
8. p-mesa
9. Open contacts
10. Planarization
11. Plated gold

Figure. 6 Schematic illustrations of the fabrication process flow: a) crucial steps before bonding, b) bonding at 280°C, c) fabrication of passives and SOAs, d) planarization and heat sink creation

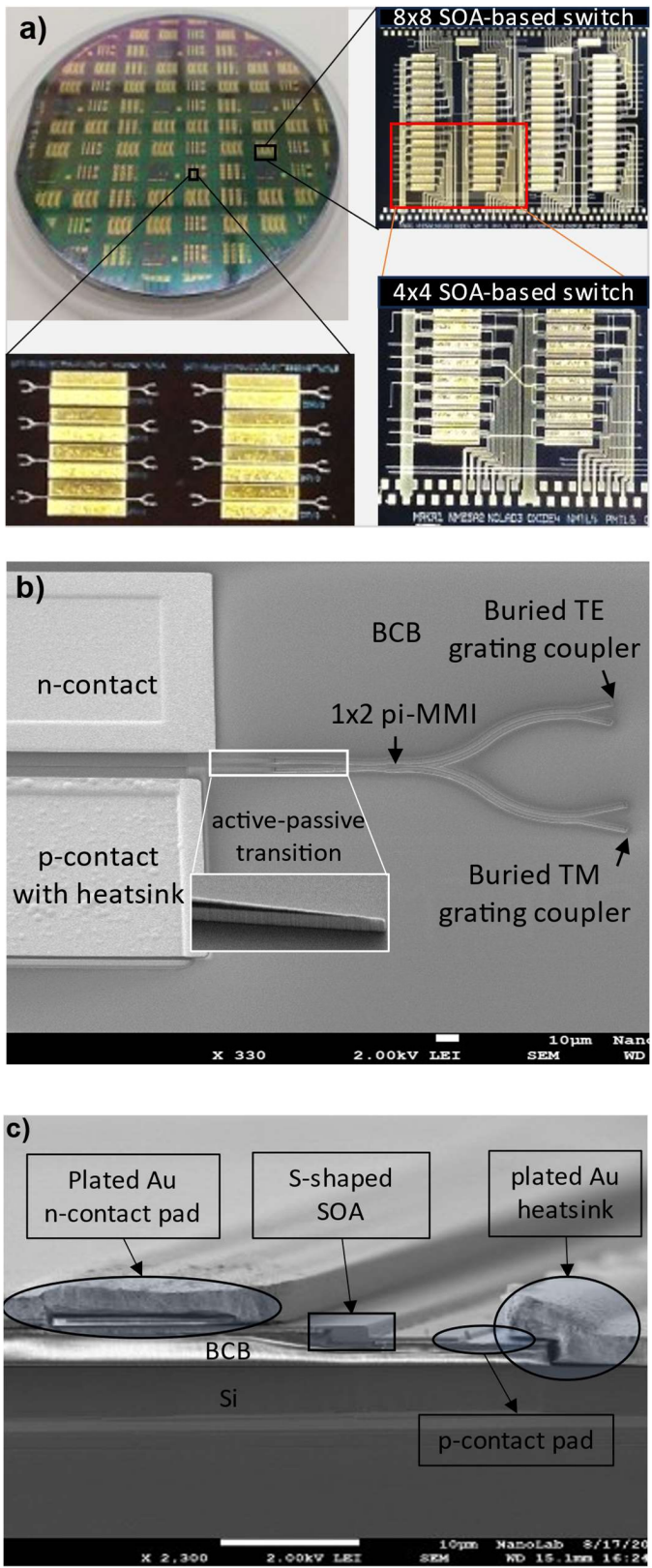

Figure 7. a) Image of the final wafer containing o-band active and passive building blocks, and optical switch circuits. b) top-view SEM image of an SOA structure used in this study c) SEM cross section of the fabricated SOA with heat sink.

5. The InP and Si wafers are prepared for bonding by depositing and outgassing 50 nm $SiO_2$ for adhesion. BCB is then spin-coated on the InP wafer.
6. The wafers are then aligned and bonded at 280°C for an hour under a vacuum to bake the BCB thoroughly. The final post-bonding BCB thickness is around 2μm with high uniformity [25] (Fig. 6b)
7. Following bonding, the processes proceed with four consecutive semiconductor dry etching steps. These steps are crucial for defining gratings, the waveguide, the final active-passive taper, and the p-mesa, which serves as the additional sidewall of the S-shaped SOA, and for creating an opening for the contacts. An efficient heat sink requires dry etching to reveal contacts proximate to the mesa sidewall, justifying the final selection. (Fig. 6c)
8. The topography of the SOA and passives is then planarized by depositing 50 nm PECVD SiO2 followed by 1μm-thick BCB layer.
9. A photolithography technique is utilized to open the planarization layer, and $CHF_3/O_2$ plasma is employed for dry etching, facilitating access to p- and n- n-contact pads from the opposite side.
10. The $SiO_2$ and BCB layers used for bonding are then removed from the p-side contact to reveal the Si for heat sink fabrication using a photolithography approach. A BCB sloped sidewall is achieved by photoresist reflow during RIE etching.
11. The final metallization for heat sinking and circuit-level pad definition starts with 45° angled Ti/Au seed layer evaporation to cover all topographic features. Patterns where Au is to be plated are patterned using photolithography. Subsequently, Au is electroplated until it forms a layer with a targeted thickness of 5 μm. The seed Au undergoes wet etching in Potassium cyanide (KCN), followed by etching of Ti using di-Ammonium hydrogen phosphate $(NH_4)_2HPO_4$. (Fig. 6d)

The final wafer, top- and cross-sectional views of the SOA are shown in Fig. 7a, 7b and 7c, respectively. The BCB thickness below the p-contact pad is around 1.8 μm. The achieved Au thickness after seed layer removal is in the 4-4.5 μm range. In Fig 7c. Due to its inherent ductility, the Au layer is slightly peeled off near the edge, which is merely an artifact due to cleaving. The plated Au layer connects well with the initial contact pads and reduces the diode resistance by 2-3 Ω from the original 10 ohms. This is crucial for driving the SOA at low power, generating less heat.

## IV. EXPERIMENT RESULTS

This section aims to find the SOA key performances. First, an experimental setup for the characterization of the SOA device is explained, followed by an explanation of methods and procedures for measuring the SOA metrics. Comments on the measured results are made in detail.

### A. Experiment setup and procedures

The fabricated device undergoes characterization to evaluate

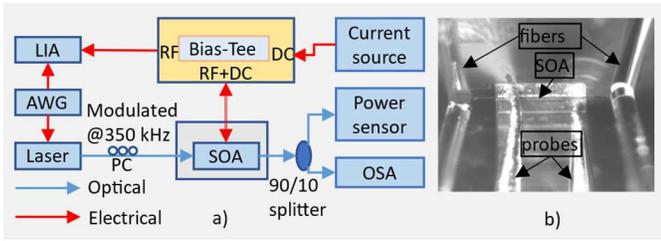

Figure 8. a) Experimental setup to measure SOA transparency current, SOA gain and SOA output power at different current and wavelength b) Measurement setup view as it is seen from microscope

gain, polarization sensitivity and NF. Fig. 8a displays the experimental configuration used to characterize the optical amplifier. The wafer with the device under test (DUT) is mounted on a copper mount with temperature control, which is stabilized to 10 °C. In this arrangement, the np junction of the SOA interfaces with the Bias-TEE terminal marked RF+DC, which links to the Lock-in Amplifier (LIA, Stanford SR865A) via its RF port and connects to the current source (Thorlabs pro8000) through its DC port. The optical amplifier receives forward bias through the Bias Tee, and concurrently, the absorption current is fed to the LIA through the RF-designated port of the Bias Tee. Modulated (350 kHz) TE/TM polarized light from a tunable laser source (Santec TLS-550) is introduced into the amplifier through a polarization controller (PC), ensuring the optimal polarization of the input light to TE/TM. The output is routed to an Optical Spectrum Analyzer (OSA) and a power meter (Keysight) through a splitter, enabling the recording of both the output optical power and power spectra. The LIA interfaced with the SOA gate facilitates assessing the response of SOA to the input optical signal. It enhances the precision of the measurements by filtering out extraneous signals and reducing measurement noise through locked-in detection. Fig. 8b shows the Device Under Test (DUT) under an optical microscope, revealing the setup where the electrical probe is aligned with the contact pads and the optical input/output is fiber-coupled.

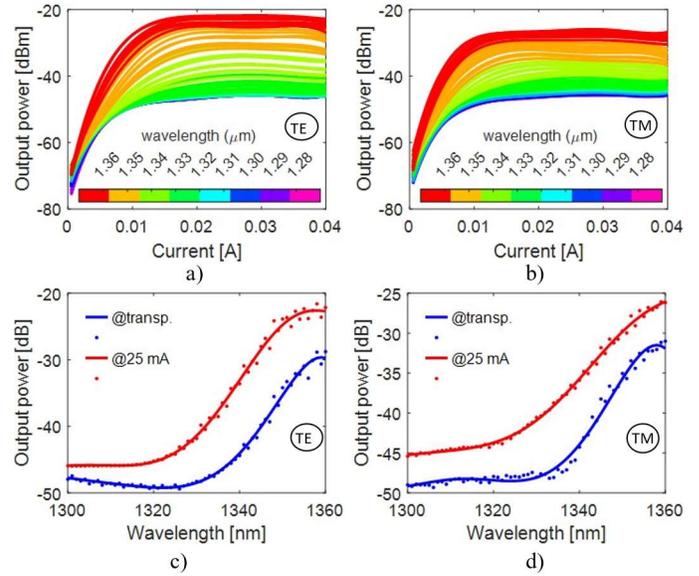

Figure. 10. a) output power as function of current for wavelength ranging from 1280 nm to 1360 nm for TE b) output power as function of current for wavelength ranging from 1280 nm to 1360 nm for TM c) output power at transparency current and at 25 mA as function of wavelength for TE d) output power at transparency current and at 25 mA as function of wavelength for TM (dots represent measurement and lines represent fitted curve).

### B. Gain and PDG measurement

The initial step in gain measurement involves identifying the transparency current. This is achieved by measuring the current within the gate of the SOA, which arises due to the interaction between the SOA and the modulated optical signal. The electrical signal measured in the SOA correlates to its absorption when the bias applied to the device is below the transparency point and to gain when the device has a bias above the transparency point [26], [27]. Therefore, the electrical signal, as recorded by the LIA, reaches a minimum when the SOA is at its transparency, representing the transition of the SOA from a state of absorption to providing net gain.

Fig. 9 shows an example of transparency current measurement results for a 500 μm long SOA as a function of current for different wavelengths. The SOA gate voltage as a function of bias current at different wavelengths is shown in

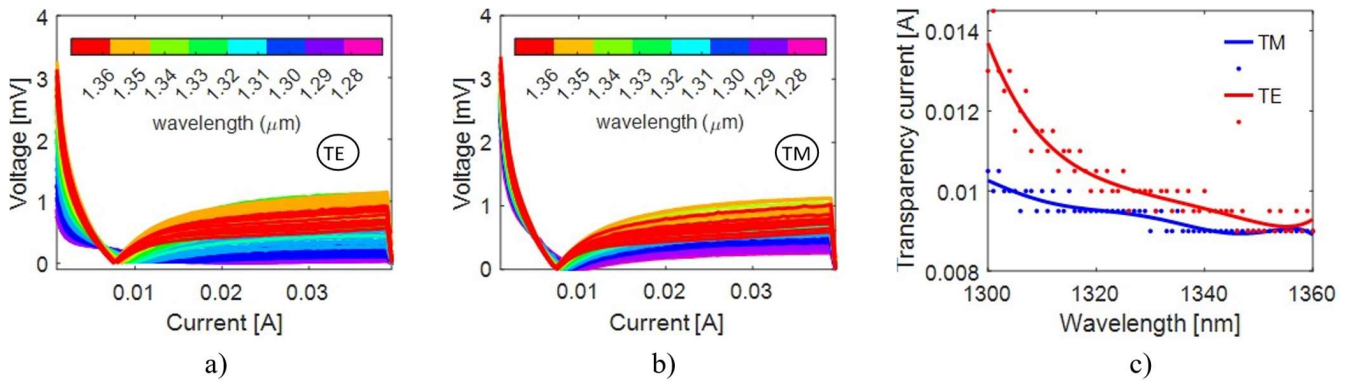

Figure 9. a) Absorption voltage as function of bias current for wavelength ranging from 1280 nm to 1360 nm for TE b) Absorption voltage as function of bias current for wavelength ranging from 1280 nm to 1360 nm for TM c) Transparency current as function of wavelength for wavelength ranging from 1300 nm to 1360 nm (dotted represent measurement and lines represent fitted curve)

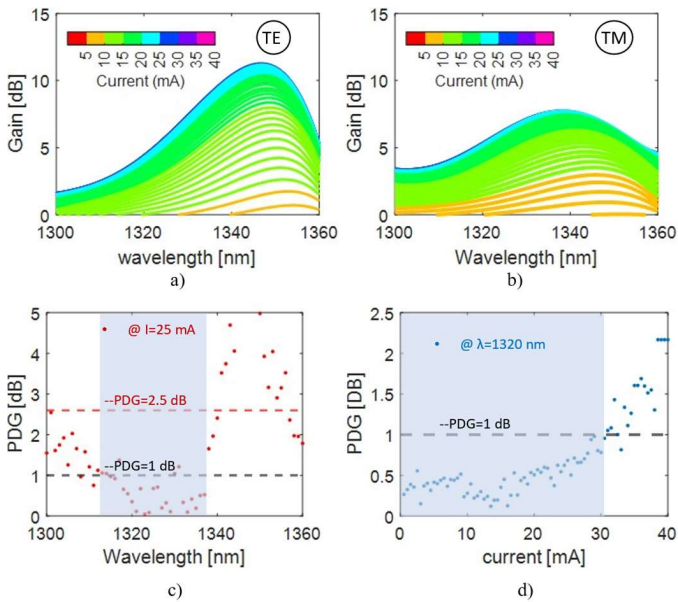

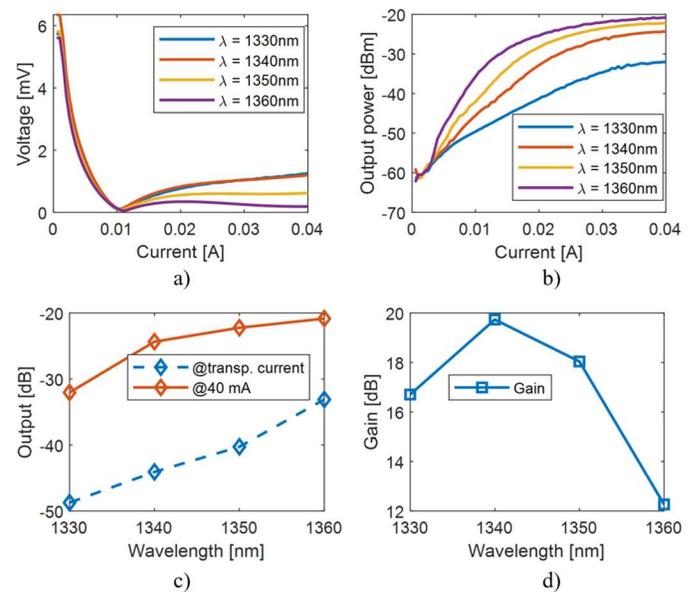

Figure 11. a) SOA gain as function of wavelength for current from 5 mA to 40 mA for TE b) SOA gain as function of wavelength for current from 5 mA to 40 mA for TM c) PDG as function of wavelength at 25 mA bias current. d) PDG as function of current at wavelength 1320 nm. The SOA gain is internal SOA gain which excludes the fiber-Grating coupler coupling losses.

Figure 12. Data obtained before the extra processing of cladding removal for comparison. a) Transparency current b) output power versus current c) output at transparency and 4kA/cm² versus wavelength. d) gain at 4kA/cm²

Fig. 9a and Fig. 9b for TE and TM, respectively. The color bar shows different wavelength ranges. The lowest voltage in the plot shows the transparency points for different wavelengths. Fig. 9c summarizes the transparency currents for different wavelengths for both TE and TM polarizations. The transparency currents for wavelengths above 1330 nm for TE and TM are less than 11 mA (1.1 kA/cm²). The transparency current for TM slightly increases as wavelength increases. On the contrary, TE experiences a significant increase in transparency current at lower wavelengths due to bandgap. Additionally, the transparency for TM is lower than TE due to bandgap shrinkage due to tensile strain.

Fig. 10a and Fig. 10b show the measured output power as a function of current for wavelength ranging from 1280 nm to 1360 nm with an on-chip input power of -10 dBm. The coupling interface for one side has 19 dB loss (3.5 dB loss from MMI, 0.4 dB loss for waveguide bend, and 15 dB loss for grating coupler) for TE and TM. The output increases for currents up to 25 mA and thus saturates for currents above 25 mA. Output at transparency and output at a certain bias current of interest can be compared to find the gain. For instance, Fig. 10c and Fig. 10d show output power at transparency current and at 25 mA as a function of wavelength for TE and TM, respectively, from which we can quickly find the gain at 25 mA.

Fig. 11a and Fig. 11b display the gain curves, plotted against wavelength, for TE and TM modes, respectively. The results are for an SOA of length 500 μm with an injected current spanning from 0.5 kA/cm² to 4 kA/cm². Measurements reveal a peak gain of 11.5 dB at 1345 nm for TE and 8 dB at 1340 nm for TM. The wavelengths at peak gain for TE and TM are 1345 nm and 1340 nm, respectively, showing only a 5 nm and 10 nm shift from the PL peak. The ~~predicted~~ 40 nm gain peak shift ~~from~~ predicted by the design due to band filling was not observed in experiment result due to several factors. The design accounted for ideal current injection efficiency, a strain coming only from the layer stack and an ideal heat sink. However, the measurement result shows much less blue shift instead, which is due to ~~, with~~ thermal effects from heat trapped at SOA sidewalls, ~~prevailing~~ over band filling effects. Additionally, strain from SOA side wall and stress from membrane bonding process, which affects gain shape has to be quantified and included. Future studies should include all of the factors. Fig. 11c represents the polarization-dependent gain (PDG), computed as GTE-GTM as in equation 1, plotted against the wavelength. The minimum PDG occurs between 1317 nm and 1337 nm wavelengths, yielding a PDG of less than 1 dB in a 25 nm bandwidth. The maximum gain in this low PDG spectral range is 8.5 dB and 7.5 dB for TE and TM polarizations, respectively (at a wavelength of 1337 nm). Operating within this range allows for considering the SOA as polarization insensitive. Polarization sensitivity stays below 2.5 dB for most wavelengths. The highest PDG occurs at the TE gain peak, where the gain for TE surpasses that for TM. Fig. 11d illustrates the PDG variation with a current up to 40 mA, measured at a fixed wavelength of 1320 nm. The PDG stays below 1 dB for currents up to 30 mA. Observations indicate that beyond 30 mA, the gain saturates and PDG increases. This is due to differential material gain under the tandem influence of gain saturation and heating which affected the polarizations differently. ~~thermal effects impact the SOA, causing a correlated increase in PDG due to heating~~.

We recorded all the gain and PDG measurements discussed above after we removed the planarization layer, the BCB, intending to improve the grating coupler coupling efficiency. Since no one had assessed the grating couplers (GC) before the run, we discovered that the peak of the GC occurs at a higher wavelength than the peak for the SOA spectrum, leading to a higher loss than anticipated. Although the removal of BCB did shift the GC transmission peak to the left, reducing insertion loss, this change altered the contact behavior of the SOA. This alteration is due to the degradation of contacts from plasma etching, modifying the SOA response at higher currents. There can also be a possibility that this process could lead to a leakage path between contacts, however comparison of IV curves before and after removing BCB do not suggest an additional leakage path–. Fig. 12 shows the SOA sample response we recorded before removing the BCB, where we could measure gain as high as 19 dB at 4kA/cm². After BCB removal, the output increases only for currents up to 2.5 kA/cm², and for currents above 2.5 kA/cm², the SOA output increases minimally.

The low output power at low wavelengths is explained by the overlap of two diminishing transfer functions as both the grating coupler efficiency and SOA gain are decreasing. Especially, the grating coupling sharply decreases at wavelengths far away from its peak wavelength. additionally, the SOA gain saturation and nonlinear properties at high input optical power has not been investigated in this work due to limitation to the maximum power we were able to couple to the chip due to the lossy on wafer vertical fiber to chip coupling. Future chip scale testing should incorporate the membrane platform compatible spot size converters to achieve high efficiency fiber to chip edge coupling.

### C. Amplifier Noise Characteristics

Noise performance is another critical metric for SOA, significantly affecting the integrity of the data signals traversing the device. The key performance indicators (KPI) related to the amplifier noise includes Amplified Spontaneous Emission (ASE), NF, and OSNR. Fig. 13a and 13b depict the noise spectrum of the SOA at varying bias currents and the OSNR at 1360 nm, respectively. The OSNR exceeds 40 dB in 1 nm resolution, suitable for applications such as optical switching where high OSNR is required. NF is calculated using equation 3, from the gain and the recorded noise spectra at the SOA output. This is calculated only for signal ASE beat noise and does not include shot noise and high order noise The calculation of NF involves using Equation 3 to determine it from the gain and the recorded noise spectra at the SOA output [28], [29]:

$$NF(\lambda) = 10 \log \frac{2\rho_{ASE}\lambda}{Ghc}, \quad\quad 3$$

where $\rho_{ASE}$ is the noise power spectral density (PSD) at the output of the SOA at the desired wavelength, $\lambda$ wavelength in m, $G$ is a single pass gain, $h$ is Planck's constant, $c$ is the light speed in a vacuum, and all parameters are given in SI unit in linear scale. Fig. 13c shows the NF as a function of wavelength. The lowest NF at 1360 nm is determined to be 6.5 dB. This shows that the SOA has good noise performance with low noise floor, good OSNR, and low NF. The fluctuation in the NF is because of the effect of the wavelength dependence of gain.

Additionally, the SOA supplies ample gain at a minimal bias current of 25 mA and a contact voltage of 1.8 V, leading to reduced power dissipation, demonstrating its appropriateness for low noise and energy-efficient dense photonic integrated circuits. For comparison, an SOA-based optical switch, developed in a generic InP platform [30], integrates 80 SOAs on the chip, operating with a bias current of 70 mA at 1.5 V, and consumes over double the power required as compared to the SOA discussed in this work. Moreover, employing polarization diversity can yield another twofold improvement in energy efficiency compared to an SOA optimized exclusively for TE.

### V. CONCLUSION

This study conceptualizes, develops, and examines a new, low PDG SOA working in the O-band. The device peak gain is 11.5 dB at 1345 nm wavelength and at 15 25 mA driving current, while the gain peak within the low PDG range is 8.5 dB at 1337 nm and . PDG of less than 1 dB over 25 nm bandwidth centered at 1325 nm wavelength. The Lowest NF which is at device gain peak is 6.5 dB. The observed minimal PDG and NF and significant gain at low driving currents highlight its potential suitability in photonic integrated circuits that need amplifier co-integration.

While this is an encouraging initial effort to develop an O-band PI SOA on the IMOS platform, there are opportunities for enhancement. First, the current model focuses on PL at 1350 nm, expecting a 40 nm blue shift because of band filling at operating currents. Nonetheless, results indicate that the thermal effects are more pronounced than those of band filling, rendering the blue shift insignificant. Consequently, future designs should aim PL at 1310 nm for effective operation in the O-band. It is also observed that the residual stresses and varying stress resulting from the bonding and manufacturing processes can potentially impact the PDG of the devices. The parameters for the model were calibrated using material properties detailed in existing studies. Thus, it is necessary to adjust the material model precisely to the membrane platform to decide the accurate strain necessary to realize low PDG.

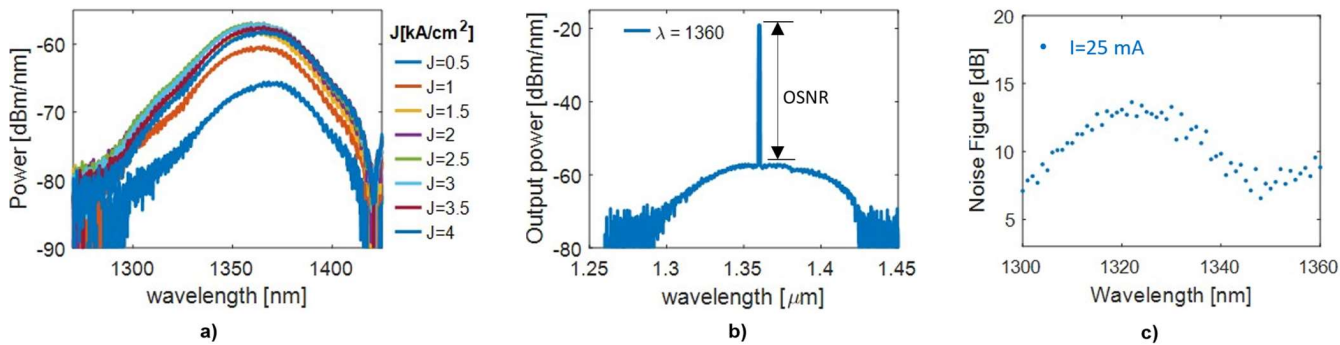

Figure 13. a) ASE spectra of the SOA for various currents b) Transmission spectrum at as function of wavelength for input at 1360 nm c) Noise Figure as function of wavelength (includes only signal ASE beat noise). The NF is calculated excluding the fiber-Grating coupler coupling losses.

Furthermore, the mismatch between the peaks of the GC transfer function and the SOA transfer function resulted in substantial coupling loss. Beyond considerations for gain material and active-passive transitions, achieving minimal loss and PI fiber-to-chip coupling is crucial to realize fully polarization-insensitive fiber-to-fiber transmission and gain.

**Desalegn Wolde Feyisa** (Student Member, IEEE) received the BS degree in electrical engineering (communication stream) from Adama Science and Technology University, Adama, Ethiopia, in 2013, the joint master's degree in photonic integrated circuits, sensors, and networks from the Scuola Superiore Sant'Anna University, Pisa, Italy, in electrical engineering from the Eindhoven University of Technology, Eindhoven, The Netherlands, in 2020. He is working toward a Ph.D. with the Electro-Optical Communication Group at Eindhoven University of Technology. His research interests include high-speed and energy-efficient optical switches.

**Salim ABDI** (Student Member, IEEE) was born in 1994 in Medea, Algeria, and obtained his bachelor's degree in materials science from Ecole National Polytechnique, Algeria. He then pursued his research master's degree in physics from Ecole Polytechnique Montreal, Canada, where he was rewarded a full scholarship from the Abdulla Al Ghurair Foundation for Education to pursue his studies. During his master's he obtained a strong background in semiconductor physics and device fabrication. His research focused on the investigation of ohmic contacts to group IV semiconductors for optoelectronic applications. Using the knowledge he acquired, he decided to continue his research career with a PhD in the field of integrated photonics. He started his project in 2021 within the Photonic Integration group at Eindhoven University of Technology, Netherlands, on the co-integration of InP electronics with InP membrane photonics for next generation transceivers from which some of the results are presented in this work.

**René van Veldhoven** (Senior Process Engineer Epitaxy) received a BEng degree in Chemical Technology in 1989. Worked on epitaxy of ferroelectric materials and silicon process integration at Philips Research from 1990-1998. After Philips Research he started working as a process engineer at JDS Uniphase, a company producing integrated photonic devices, where he was responsible for Metal Organic Vapor Phase Epitaxy. Since 2002 until now he has been working as a process engineer in the applied physics department in the Nanolab group. In this group he is responsible for several MOVPE and MBE set-ups and involved in the growth of nanowires and III-V materials used in photonic integration.

**Nicola Calabretta** (Member, IEEE) received the BS and MS degrees in telecommunications engineering from Politecnico di Torino, Turin, Italy, in 1995 and 1999, respectively, and the Ph.D. degree from the COBRA Research Institute, Eindhoven University of Technology, Eindhoven, The Netherlands, in 2004. From 2004 to 2007, he was a Researcher with Scuola Superiore Sant'Anna University, Pisa, Italy. He is currently an Associate Professor with the Institute of Photonic Integration, Eindhoven University of Technology. He has co-authored more than 490 papers published in international journals and conferences and holds five patents. His research interests include smart optical data centers and telecom optical networks, optical processing, optical packet switching, and semiconductor optical amplifier based optical switching and processing.

**Yuqing Jiao** (Senior Member, IEEE) received dual Ph.D. degrees from the Eindhoven University of Technology, The Netherlands, as well from Zhejiang University, China, in 2013. Since then, he has been focusing on the research of InP-based nano photonics at the Institute of Photonic Integration (IPI, former COBRA Research Institute) within Eindhoven University of Technology. He is currently an Associate Professor with the institute. He has coauthored more than 60 international journal publications and 150 conference articles and holds four patents. His research interests include integrated nano photonics and heterogeneous integration technologies. He currently serves as a Board Member for the IEEE Photonics Society Benelux Chapter.

**Ripalta Stabile** (Senior Member, IEEE) received the master's degree in electrical engineering from the Politecnico of Bari, Bari, Italy, in 2004, and the Ph.D. degree in nanoscience from National Nanotechnology Laboratory, Lecce, Italy, in 2008. In 2009, she moved to COBRA Research Institute, Eindhoven University of Technology, Eindhoven, The Netherlands, where she was appointed as a Postdoctoral Researcher. She is currently an Associate Professor with the Institute of Photonic Integration (formerly COBRA), Eindhoven University of Technology. Her research interests include design, simulation, nanofabrication, and characterization of organic and inorganic photonic devices. She is an Expert in Indium Phosphide (InP) large-scale photonic integrated circuits based on semiconductor-optical-amplifiers design for next generation optical networks and on-chip photonic neural networks.